\documentclass[twocolumn,aps,floatfix,nofootinbib]{revtex4-1}
\usepackage{amsmath}
\usepackage{epsfig}
\usepackage{slashed}
\begin{document}
\title{Anomalous Dimensions and the Renormalizability of the Four-Fermion Interaction}
\author{Philip D. Mannheim\\
Department of Physics, University of Connecticut, Storrs, CT 06269, USA\\
email: philip.mannheim@uconn.edu}
\date{May 10, 2017}

\begin{abstract}
We show that when the dynamical dimension of the $\bar{\psi}\psi$ operator is reduced from three to two in a fermion electrodynamics with scaling, a $g(\bar{\psi}\psi)^2+g(\bar{\psi}i\gamma^5\psi)^2$ four-fermion interaction which is dressed by this electrodynamics becomes renormalizable. In the fermion-antifermion scattering amplitude every term in an expansion to arbitrary order in $g$  is found to diverge as just a single ultraviolet logarithm (i.e. no log squared or higher), and is thus made finite by a single subtraction. While not necessary for renormalizability per se, the reduction in the dimension of $\bar{\psi}\psi$ to two leads to dynamical chiral symmetry breaking in the infrared, with the needed subtraction then automatically being provided by the theory itself through the symmetry breaking mechanism, with there then being no need to introduce the subtraction by hand. Since the vector and axial vector currents are conserved, they do not acquire any anomalous dimension, with the four-fermion  $(\bar{\psi}\gamma^{\mu}\psi)^2$ and $(\bar{\psi}\gamma^{\mu}\gamma^5\psi)^2$ interactions instead having to be controlled by the standard Higgs mechanism.
\end{abstract}
\maketitle

\section{QED and the Four-Fermion Theory}

In this paper we study the ultraviolet structure of a fermion gauge theory (generically quantum electrodynamics (QED)) coupled to a scalar plus pseudoscalar four-fermion (FF) theory, with action $I^m_{\rm QED}+I_{\rm FF}$, where
\begin{eqnarray}
I^m_{\rm QED}&=&\int d^4x\bigg{[}-\frac{1}{4}F_{\mu\nu}F^{\mu\nu}+\bar{\psi}\gamma^{\mu}(i\partial_{\mu}-eA_{\mu})\psi 
\nonumber\\
&-&m\bar{\psi}\psi\bigg{]},
\nonumber\\
 I_{\rm FF}&=&\int d^4x\bigg{[}-\frac{g}{2}[\bar{\psi}\psi]^2-\frac{g}{2}[\bar{\psi}i\gamma^5\psi]^2\bigg{]}.
\label{A1a}
\end{eqnarray}
In our study we impose only one requirement, namely that the effect of QED photon exchanges is to dress the fermion bilinear $\bar{\psi}\psi$ so that its dimension $d_{\theta}(\alpha)$ is dynamically reduced from three to two. With such a dressing we are then able to show that the four-fermion interaction becomes renormalizable to all orders in the four-fermion coupling constant $g$. To be specific, we note that in the generalized Landau gauge the asymptotic renormalized inverse massive fermion propagator $\tilde{S}_m^{-1}(p)$ obeys \cite{Adler1971} the Callan-Symanzik equation

\begin{eqnarray}
\left[m\frac{\partial}{m}+\beta(\alpha)\frac{\partial}{\partial \alpha}\right]\tilde{S}_m^{-1}(p)=-m[1-\gamma_{\theta}(\alpha)]\tilde{\Gamma}^m_{\rm S}(p,p,0),~~~
\label{A2a}
\end{eqnarray}
where $\tilde{\Gamma}^m_{\rm S}(p,p,0)$ is the renormalized Green's function associated with the insertion of a zero-momentum composite operator $\theta=\bar{\psi}\psi$ into the inverse massive fermion propagator. Asymptotic scaling is achievable if $\beta(\alpha)=0$, with one then asymptotically having 
\begin{eqnarray}
\tilde{S}_m^{-1}(p)&=& \slashed{p}-m\left(\frac{-p^2-i\epsilon}{m^2}\right)^{\frac{\gamma_{\theta}(\alpha)}{2}}+i\epsilon,
\nonumber\\
\tilde{\Gamma}^m_{\rm S}(p,p,0)&=&\left(\frac{-p^2-i\epsilon}{m^2}\right)^{\frac{\gamma_{\theta}(\alpha)}{2}},
\label{A3a}
\end{eqnarray}
where $\gamma_{\theta}(\alpha)$ is the anomalous part of $d_{\theta}(\alpha)$ as defined as $d_{\theta}(\alpha)=3+\gamma_{\theta}(\alpha)$. The insertion of the $\tilde{\Gamma}^m_{\rm S}$ vertex into the four-fermion Green's functions will then soften them sufficiently in the ultraviolet to make them renormalizable if $d_{\theta}(\alpha)$ is reduced from three to two. The possibility that $d_{\theta}(\alpha)$ would be equal to two has been much discussed in the literature (see e.g.  \cite{Miransky1993,Mannheim2015,Mannheim2016a} and references therein). Moreover the $d_{\theta}(\alpha)=2$ condition has been discussed not just in the Abelian theory that we study here but even in the non-Abelian case as well, where it is considered to be a condition for dynamical symmetry breaking in walking technicolor scenarios (see e.g. \cite{Cohen1989} and references therein).

QED studies that involve photons that are not dressed at all (quenched photons) can be associated with a $\beta(\alpha)$ that is zero identically for any value of $\alpha$, with (\ref{A3a}) then holding for any value of $\alpha$. If one studies the theory in the ladder (planar graph) approximation with such quenched photons one precisely obtains our required $d_{\theta}(\alpha)=2$ if the coupling constant $\alpha$ takes the value $\alpha=\pi/3$ \cite{Maskawa1974}. One can go beyond the quenched ladder approximation and include quenched non-planar graphs as well, and to all orders in quenched photon exchanges one again finds \cite{Johnson1964} asymptotic scaling. Moreover, this all-order result holds for all possible values of $\alpha$ weak or strong, and since $d_{\theta}(\alpha)$ is a continuous function of $\alpha$ one can in principle choose a large enough $\alpha$ so that $d_{\theta}(\alpha)$ is reduced to two (or one can make the theory even more convergent if $d_{\theta}(\alpha)$ is reduced to below two). If one wants to include non-quenched photons ($\beta(\alpha)$ then not zero identically) the scaling result of  \cite{Johnson1964} will continue to hold \cite{Johnson1967} if, as was subsequently reformulated in \cite{Adler1971}, $\beta(\alpha)$ has a zero away from the origin. With the beta function actually being zero identically if the photon is quenched,  in all these cases the beta function vanishes, and thus in all cases one has asymptotic scaling with anomalous dimensions. 

Regardless of whether or not the QED beta function might actually vanish away from the origin, the results we present in this paper can be understood as the statement that when quenched photon exchanges cause $d_{\theta}(\alpha)$ to be reduced from three to two, the four-fermion theory vertices are softened enough to make the four-fermion theory renormalizable. In fact, suppose one has a set of quenched QED exchange graphs for which the four-fermion interaction is made renormalizable. Then including non-quenched QED exchange graphs will not change this since from this point on the QED and four-fermion sectors are both power-counting renormalizable.\footnote{Starting with \cite{Mannheim1974,Mannheim1975,Mannheim1978} and \cite{Leung1986} it had been suggested that the condition $d_{\theta}(\alpha)=2$ could lead to the renormalizability of the four-fermion interaction. In this paper this is explicitly shown to be the case to all orders in the four-fermion interaction.}

The actual analysis that we present in this paper is divided into two parts, a study of QED plus a four-fermion interaction when the chiral symmetry vacuum is unique, and a study of the same theory when there is dynamical symmetry breaking and the vacuum is degenerate, a symmetry breaking that is actually caused by the very same $d_{\theta}(\alpha)=2$ condition \cite{Mannheim1974,Mannheim1975,Mannheim1978}. Since dynamical symmetry breaking is an infrared effect, the ultraviolet behavior of the theory is the same in both the unique and degenerate vacuum cases, and to establish renormalizability per se it suffices to study the unbroken vacuum case alone. However, when we do study the broken vacuum case we not only find renormalizability we find finiteness, with the symmetry breaking procedure itself automatically generating the needed counterterms for us.

\section{All-Order Renormalizability of the Four-Fermion Interaction}

Since the ultraviolet behavior of QED is not sensitive to mass, to establish the renormalizability of the four-fermion interaction we can drop the mass term in (\ref{A1a}) and study the chirally-symmetric $I^0_{\rm QED}+I_{\rm FF}$, where
\begin{eqnarray}
&&I^0_{\rm QED}=\int d^4x\bigg{[}-\frac{1}{4}F_{\mu\nu}F^{\mu\nu}+\bar{\psi}\gamma^{\mu}(i\partial_{\mu}-eA_{\mu})\psi \bigg{]},~
\nonumber\\
&&I_{\rm FF}=\int d^4x\bigg{[}-\frac{g}{2}[\bar{\psi}\psi]^2-\frac{g}{2}[\bar{\psi}i\gamma^5\psi]^2\bigg{]}.
\label{A4a}
\end{eqnarray}
The Green's functions that are needed for the scattering generated by $I_{\rm FF}$ are those that involve $\bar{\psi}\psi$ and $\bar{\psi}i\gamma^5\psi$. The utility of studying the massless fermion limit is that in this limit what had been only asymptotic scaling now becomes scaling at all momenta, both asymptotic and non-asymptotic. In this limit the exact two-point $\bar{\psi}\psi$ Green's function is given by (see e.g. \cite{Mannheim1975})
\begin{eqnarray}
\langle \Omega_0|T(\bar{\psi}(x)\psi(x)\bar{\psi}(0)\psi(0))|\Omega_0\rangle
=\mu^{-2\gamma_{\theta}(\alpha)}(x^2)^{-d_{\theta}(\alpha)},~~
\label{A5a}
\end{eqnarray}
where $\mu$ is an off-shell subtraction point that is needed for a massless theory. Fourier transforming then gives 
\begin{eqnarray}
\Pi_{\rm S}^{0}(q^2)&=&-i\int \frac{d^4p}{(2\pi)^4}{\rm Tr}\bigg{[}\left(\frac{(-p^2)}{\mu^2}\frac{(-(p+q)^2)}{\mu^2}\right)^{\frac{\gamma_{\theta}(\alpha)}{4}}
\nonumber\\
&\times&\frac {1}{\slashed{ p}}\left(\frac{(-p^2)}{\mu^2}\frac{(-(p+q)^2)}{\mu^2}\right)^{\frac{\gamma_{\theta}(\alpha)}{4}}\frac {1}{\slashed{ p} +\slashed {q}}\bigg{]}.
\label{A6a}
\end{eqnarray}
On making a Dyson-Wick contraction of the fields on the left-hand side of (\ref{A5a}), we obtain 
\begin{eqnarray}
\Pi_{\rm S}^0(q^2)&=&-i\int \frac{d^4p}{(2\pi)^4}{\rm Tr}\bigg{[}\tilde{S}_0(p)\tilde{\Gamma}^0_{\rm S}(p,p+q,q)
\nonumber\\
&\times&\tilde{S}_0(p+q)\tilde{\Gamma}^0_{\rm S}(p+q,p,-q)\bigg{]},
\label{A7a}
\end{eqnarray}
where $\tilde{S}_0(p)=1/\slashed{ p}$, and $\tilde{\Gamma}^0_{\rm S}(p,p+q,q)$ is the Green's function associated with the insertion of the $\bar{\psi}\psi$ vertex with momentum $q_{\mu}$ into the inverse massless fermion propagator. Comparing (\ref{A6a}) and (\ref{A7a}), and in parallel to (\ref{A3a}), we obtain
\begin{eqnarray}
\tilde{\Gamma}^0_{\rm S}(p,p+q,q)&=&\left[\frac{(-p^2)}{\mu^2}\frac{(-(p+q)^2)}{\mu^2}\right]^{\frac{\gamma_{\theta}(\alpha)}{4}}
\nonumber\\
&=&\tilde{\Gamma}^0_{\rm S}(p+q,p,-q),
\label{A8a}
\end{eqnarray}
a relation that we will need in the following.

On  using
\begin{eqnarray}
\Pi_{j=1}^nA_j^{-\lambda_j}=\frac{\Gamma(\sum\lambda_j)}{\Pi_j\Gamma(\lambda_j)}\int \Pi_id\alpha_i\frac{\delta(1-\sum\alpha_i)\Pi_i\alpha_i^{\lambda_i-1}}{(\sum\alpha_iA_i)^{(\sum\lambda_i)}},~~
\label{A9a}
\end{eqnarray}
we can evaluate $\Pi_{\rm S}^0(q^2)$ as given in (\ref{A7a}) analytically. For spacelike $q^2$, the region of interest for renormalization, we can Wick rotate, and with $\gamma_{\theta}(\alpha)=-1$ and $q^2_E=-q^2$ obtain
\begin{eqnarray}
\Pi_{\rm S}^{0}(q^2)&=&-4i\int \frac{d^4p}{(2\pi)^4}\frac{(-\mu^2)p.(p+q)}{[p^2(p+q)^2]^{3/2}}
\nonumber\\
&=&-\frac{2\mu^2}{\pi^3}\int _0^{\Lambda^2}du^2 u^2\int _0^1d\alpha_1\frac{A^{1/2}(u^2-Aq_E^2)}{(u^2+Aq_E^2)^3},~~~~
\label{A10a}
\end{eqnarray}
where $u_{\mu}=p_{\mu}+\alpha_1 q_{\mu}$, and $A=\alpha_1(\alpha_1-1)$. With (\ref{A10a}) entailing that $\Pi_{\rm S}^{0}(q^2)/\mu^2$ is dimensionless, the leading divergence of $\Pi_{\rm S}^{0}(q^2)$ is just a single logarithm, viz. 
\begin{eqnarray}
\Pi_{\rm S}^{0}(q^2)
=-\frac{\mu^2}{4\pi^2}{\rm ln}\left(\frac{\Lambda^2}{-q^2}\right).
\label{A11a}
\end{eqnarray}
With a point-coupled $\tilde{\Gamma}^0_{\rm S}(p,p+q,q)=1$ vertex (viz. $d_{\theta}(\alpha)=3$) leading to a quadratic divergence in $\Pi_{\rm S}^{0}(q^2)$ (the familiar Nambu-Jona-Lasinio model situation \cite{Nambu1961}), we see that dressing $\tilde{\Gamma}^0_{\rm S}(p,p+q,q)$ to all orders in $\alpha$ as in (\ref{A8a})  leads to a $\Pi_{\rm S}^{0}(q^2)$ that is only log divergent when $d_{\theta}(\alpha)=2$. Because of the chiral symmetry of the massless theory, exactly the same analysis holds for the pseudoscalar 
\begin{eqnarray}
\Pi_{\rm P}^0(q^2)&=&-i\int \frac{d^4p}{(2\pi)^4}{\rm Tr}\bigg{[}\tilde{S}_0(p)\tilde{\Gamma}^0_{\rm S}(p,p+q,q)i\gamma^5
\nonumber\\
&\times&\tilde{S}_0(p+q)\tilde{\Gamma}^0_{\rm S}(p+q,p,-q)i\gamma^5\bigg{]},
\label{A12a}
\end{eqnarray}
with $\Pi_{\rm P}^0(q^2)$ being equal to $\Pi_{\rm S}^0(q^2)$. Both have the same asymptotic behavior in the ultraviolet, something  that will continue to hold after the chiral symmetry of $I^0_{\rm QED}+I_{\rm FF}$ is broken in the infrared.

On now coupling $I^0_{\rm QED}$ to $I_{\rm FF}$, to lowest order in $g$ in the Bethe-Salpeter kernels for the scalar and the pseudoscalar channel fermion-antifermion scattering amplitudes $T^0_{\rm S}(q^2)$ and $T^0_{\rm P}(q^2)$, we obtain $T^0_{\rm S}(q^2)=g+g\Pi_{\rm S}^{0}(q^2)g+....$, $T^0_{\rm P}(q^2)=g+g\Pi_{\rm P}^{0}(q^2)g+....$, viz.
\begin{eqnarray}
T^0_{\rm S}(q^2)=\frac{1}{g^{-1}-\Pi^0_{\rm S}(q^2)},~T^0_{\rm P}(q^2)=\frac{1}{g^{-1}-\Pi^0_{\rm P}(q^2)}.~~
\label{A13a}
\end{eqnarray}
We can thus choose $g^{-1}$ to be a single log divergence, with $T^0_{\rm S}(q^2)$ and $T^0_{\rm P}(q^2)$ then both being finite. We will see below that the symmetry breaking procedure will naturally lead to a $g^{-1}$ with a log divergence that will precisely cancel the log divergences of $\Pi^0_{\rm S}(q^2)$ and $\Pi^0_{\rm P}(q^2)$.

\begin{figure}[htpb]
\epsfig{file=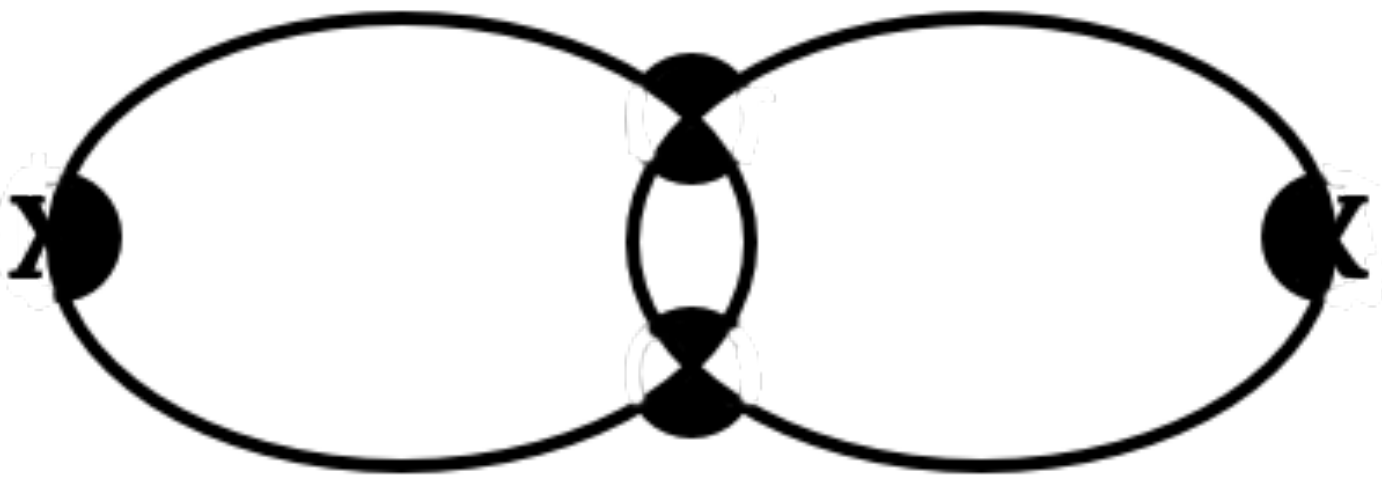,width=3.5cm,height=2.3cm}
\epsfig{file=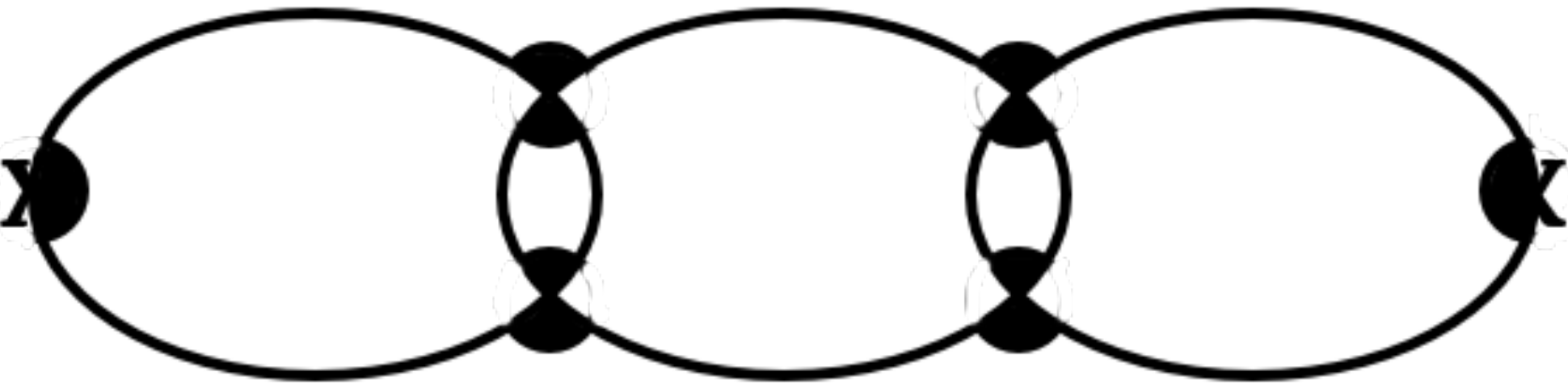,width=5.0cm,height=2.3cm}\\
\caption{Order $g^2$ and $g^4$ contributions to $\Pi^0_{\rm S}(q^2)$. The blobs denote $\tilde{\Gamma}^0_{\rm S}(p,p+q,q)$ with appropriate momenta.}
\label{higherorder}
\end{figure}

As well as dress  $\Pi^0_{\rm S}(q^2)$ and  $\Pi^0_{\rm P}(q^2)$ with QED contributions, we also need to dress them with higher order $I_{\rm FF}$ contributions. The first two such contributions are shown in Fig. (\ref{higherorder}). To determine exactly where to put the dressed $\tilde{\Gamma}^0_{\rm S}(p,p+q,q)$ vertex and its pseudoscalar analog $\tilde{\Gamma}^0_{\rm P}(p,p+q,q)$, we note that in the path integral ${\int}D[\bar{\psi}]D[\psi]D[A_{\mu}]\exp[i(I^0_{\rm QED}+I_{\rm FF})]$ we can add in dummy Gaussian integrations ${\int}D[\sigma]D[\pi]\exp\left[i\left((\sigma-g\bar{\psi}\psi)^2/2g+(\pi-g\bar{\psi}i\gamma^5\psi)^2/2g\right)\right]$. When combined with the four-fermion terms in $I_{\rm FF}$ this leads to a net contribution of the form $\int D[\sigma]D[\pi]\exp\left[i\left(-\sigma\bar{\psi}\psi+\sigma^2/2g-\pi\bar{\psi}i\gamma^5\psi +\pi^2/2g\right)\right]$, to thus effectively break up the point four-fermion interactions into $\sigma$- and $\pi$-mediated Yukawa interactions with zero-momentum, point-like propagators $1/(-m_{\sigma}^2)=g$, $1/(-m_{\pi}^2)=g$. The scalar and pseudoscalar Yukawa vertices are then dressed with $\tilde{\Gamma}^0_{\rm S}(p,p+q,q)$ and $\tilde{\Gamma}^0_{\rm P}(p,p+q,q)$. This gives the pattern of  vertex dressings exhibited in Fig. (\ref{higherorder}).

With (\ref{A8a}), at $\gamma_{\theta}(\alpha)=-1$ the $g^2$ contribution is given by 
\begin{eqnarray}
&&\Pi_{\rm S}^0(q^2,2)=-i\int \frac{d^4p}{(2\pi)^4}\frac{d^4k}{(2\pi)^4}\frac{d^4r}{(2\pi)^4}\frac{(-\mu^2)^3g^2X}{Y},~~
\nonumber\\
&&Y=[p^2(p+q)^2(p+q+k)^2(p+k)^2r^2(r+k)^2]^{3/2},~~
\nonumber\\
&&X={\rm Tr}[\slashed{p}(\slashed{p}+\slashed{q})(\slashed{p}+\slashed{q}+\slashed{k})(\slashed{p}+\slashed{k})]
{\rm Tr}[\slashed{r}(\slashed{r}+\slashed{k})].~~
\label{A14a}
\end{eqnarray}
With $\Pi_{\rm S}^0(q^2,2)/g^2\mu^6$ being dimensionless, through use of (\ref{A9a}) we find that,  up to a numerical coefficient, $\Pi_{\rm S}^0(q^2,2)$ diverges as a single logarithm of the form $\mu^6g^2{\rm ln}[\Lambda^2/(-q^2)]$. The same pattern repeats for the $g^4$ contribution $\Pi_{\rm S}^0(q^2,4)$, with each two powers of $g$ adding in two  powers of $\mu^2$, two momentum integrations, four powers of momentum in the numerator, and 12 powers of momentum in the denominator. Thus no matter how many additional terms we add in to the kernel, $\Pi_{\rm S}^0(q^2,{\rm all})=\Pi_{\rm S}^0(q^2)+\sum _{n=1}^{\infty}\Pi_{\rm S}^0(q^2,2n)$ continues to diverge as a single log. Hence, with just one subtraction $\Pi_{\rm S}^0(q^2,{\rm all})$ and $\Pi_{\rm P}^0(q^2,{\rm all})$ become ultraviolet finite. 

To explicitly implement the needed subtraction, we first rewrite $g$ as $g=G/\mu^2$ where $G$ is dimensionless, so that we can then write the divergent part of the all-order $\Pi_{\rm S}^0(q^2,{\rm all})$  in the form $-(\mu^2/4\pi^2)F(G){\rm ln}[\Lambda^2/(-q^2)]$ where $F(G)$ is power series in $G^2$ with $F(0)=1$, and can thus set  
\begin{eqnarray}
T^0_{\rm S}(q^2)=\frac{4\pi^2}{\mu^2\left(4\pi^2G^{-1}+F(G){\rm ln}[\Lambda^2/(-q^2)]\right)}.
\label{A15a}
\end{eqnarray}
We then obtain 
\begin{eqnarray}
T^0_{\rm S}(q^2)=\frac{4\pi^2}{\mu^2F(G){\rm ln}[\mu^2/(-q^2)]}
\label{A16a}
\end{eqnarray}
if we require $G$ to obey  
\begin{eqnarray}
4\pi^2G^{-1}+F(G){\rm ln}[\Lambda^2/\mu^2]=0.
\label{A17a}
\end{eqnarray}
For the typical case of $F(G)=1+G^2$ this requirement yields a leading behavior for $G$ of the form 
\begin{eqnarray}
G=-\frac{4\pi^2}{{\rm ln}[\Lambda^2/\mu^2]}+\frac{(4\pi^2)^3}{{\rm ln}^3[\Lambda^2/\mu^2]}
\label{A18a}
\end{eqnarray}
for large cutoff.\footnote{With the bare coupling constant $G$ vanishing as the inverse of a logarithm, the theory has a structure reminiscent of that found in asymptotically free theories. A connection to asymptotic freedom was also suggested in \cite{Mannheim1975} from a study of the structure of the vacuum energy density $\tilde{\epsilon}(m)$ that appears in (\ref{A25a}) below.} Consequently one can set $F(G)=1$, and obtain the completely finite   
\begin{eqnarray}
T^0_{\rm S}(q^2)=T^0_{\rm P}(q^2)=\frac{4\pi^2}{\mu^2{\rm ln}[\mu^2/(-q^2)]}.
\label{A19a}
\end{eqnarray}

Obtaining just a single log divergence is quite familiar in field theories with an underlying scale or conformal symmetry. For instance, since the vector current $j_{\mu}$ is conserved, it remains canonical under interactions if there is conformal invariance, with the all-order QED vacuum polarization then being given by $\Pi_{\mu\nu}(z)=\langle\Omega|T(j_{\mu}(z)j_{\nu}(0))|\Omega\rangle=f(\alpha)(\eta_{\mu\nu}\partial_{\alpha}\partial^{\alpha}-\partial_{\mu}\partial_{\nu})z^{-4}$ in coordinate space and by $\Pi_{\mu\nu}(q^2)=if(\alpha)(\eta_{\mu\nu}q^2-q_{\mu}q_{\nu}){\rm ln}[\Lambda^2/(-q^2)]$ in momentum space (see e.g. \cite{Mannheim1975b}), where $f(\alpha)$ is a dimensionless function of $\alpha$.

The momentum  dependence pattern found for $\Pi_{\rm S}^0(q^2,2n)$ has a parallel in the conformal gravity theory  that has been advanced (see e.g. \cite{Mannheim2006,Mannheim2012,Mannheim2015}) as a candidate alternative to standard Einstein gravity and its quantum string theory generalization.  Conformal gravity is based on the action $I _{\rm W}=-\alpha_g\int d^4x (-g)^{1/2}C_{\lambda\mu\nu\kappa} C^{\lambda\mu\nu\kappa}$ where $C_{\lambda\mu\nu\kappa}$ is the conformal Weyl tensor,  and with the gravitational coupling constant $\alpha_g$ being dimensionless, conformal gravity is power-counting renormalizable.  Specifically, with $g_{\mu\nu}$ being dimensionless, in an expansion around flat spacetime of the dimension four quantity $C_{\lambda\mu\nu\kappa} C^{\lambda\mu\nu\kappa}$ as a power series in a gravitational fluctuation $h_{\mu\nu}=g_{\mu\nu}-\eta_{\mu\nu}$, each term will contain $h_{\mu\nu}$  a specific number of times together with exactly four derivatives since it is the derivatives that carry the dimension of the $C_{\lambda\mu\nu\kappa} C^{\lambda\mu\nu\kappa}$ term. The term that is quadratic in $h_{\mu\nu}$ will thus give a $1/k^4$ propagator, and each time we work to one more order in $h_{\mu\nu}$ we add one extra $1/k^4$ propagator and a compensating factor of $k_{\lambda}k_{\mu}k_{\nu}k_{\kappa}$ in the vertex. With equal numbers of powers of $k_{\mu}$ being added in the numerator and denominator (just as in $\Pi_{\rm S}^0(q^2,2n)$), renormalizability is thereby secured.\footnote{Because the conformal gravity theory is based on fourth-order derivative equations, it had been thought that the theory would possess  negative Dirac norm ghost states and not be unitary. However, on expressly quantizing the theory and constructing the appropriate quantum Hilbert space, it was found \cite{Bender2008a,Bender2008b} that the Hamiltonian of the theory was not Hermitian but  was instead $PT$ symmetric. Thus one could not use the Dirac norm, but had to instead use the norm associated with $PT$ symmetry program of Bender and collaborators \cite{Bender2007}, and this norm was expressly found to not be negative.  In consequence conformal gravity can be regarded as a fully consistent and unitary theory of quantum gravity in its own right.}

Moreover, in a conformal theory  $T_{\mu\nu}$ is both transverse and traceless, and has canonical dimension four. In terms of the transverse-traceless projector given in \cite{Mannheim2005}, viz. $P_{\mu\nu\sigma\tau}=(1/2)(P_{\mu\sigma}P_{\nu\tau}+P_{\mu\tau}P_{\nu\sigma})-(1/3)P_{\mu\nu}P_{\sigma\tau}$ where $P_{\mu\nu}=\eta_{\mu\nu}\partial_{\alpha}\partial^{\alpha}-\partial_{\mu}\partial_{\nu}$, we can set $\langle \Omega_0|T(T_{\mu\nu}(z)T_{\sigma\tau}(0))|\Omega_0\rangle=P_{\mu\nu\sigma\tau}z^{-4}$. Just as with the photon $\Pi_{\mu\nu}(q^2)$, the Fourier transform of $P_{\mu\nu\sigma\tau}z^{-4}$ behaves as a single log.

With vector and axial vector currents both remaining canonical, four-fermion interactions of the form $\bar{\psi}\gamma_{\mu}\psi\bar{\psi}\gamma^{\mu}\psi$ and $\bar{\psi}\gamma_{\mu}\gamma^5\psi\bar{\psi}\gamma^{\mu}\gamma^5\psi$ cannot be made renormalizable by anomalous dimensions, and for them we must use the standard intermediate gauge boson Higgs mechanism. With the energy-momentum tensor $T_{\mu\nu}$ being conserved, it also remains canonical. Thus for interactions of the form $T_{\mu\nu}T^{\mu\nu}$, to achieve renormalizability we must use the conformally coupled  gravitons of conformal gravity as intermediaries. However, $I_{\rm FF}$ can be renormalized by anomalous dimensions, and we discuss now how this relates to dynamical mass generation. That there would be dynamical mass generation in a model based on  $I^0_{\rm QED}+I_{\rm FF}$ can immediately be anticipated \cite{Mannheim2015,Mannheim2016a} since in the massless vacuum $|\Omega_0\rangle$ we obtained $T^0_{\rm S}(q^2)=T^0_{\rm P}(q^2)=4\pi^2/\left[\mu^2{\rm ln}[\mu^2/(-q^2)]\right]$, to thus put tachyons in $T^0_{\rm S}(q^2)$ and $T^0_{\rm P}(q^2)$ at $q^2=-\mu^2$, and thereby render the massless vacuum unstable.

\section{Dynamical Mass Generation and Finiteness of the Four-Fermion Interaction}

Consideration of dynamical symmetry breaking in the theory based on $I^0_{\rm QED}+I_{\rm FF}$ leads us \cite{Mannheim1974,Mannheim1975,Mannheim1978} to a unique value for $\gamma_{\theta}(\alpha)$, namely $\gamma_{\theta}(\alpha)=-1$ (i.e. $d_{\theta}(\alpha)=2$), as this is the condition that the chiral vacuum break. Essentially, the more negative $\gamma_{\theta}(\alpha)$ is taken to be, the more convergent the theory becomes in the ultraviolet, and thus the more divergent it becomes in the infrared, with infrared divergences then uniquely forcing us into a long range order broken vacuum and a dynamically generated double-well potential when $\gamma_{\theta}(\alpha)=-1$. (If $0>\gamma_{\theta}(\alpha)>-1$ the potential is a single well, and if $\gamma_{\theta}(\alpha)<-1$ the potential is an unbounded upside down single well   \cite{Mannheim1974}.) 

\begin{figure}[htpb]
\begin{center}
\epsfig{file=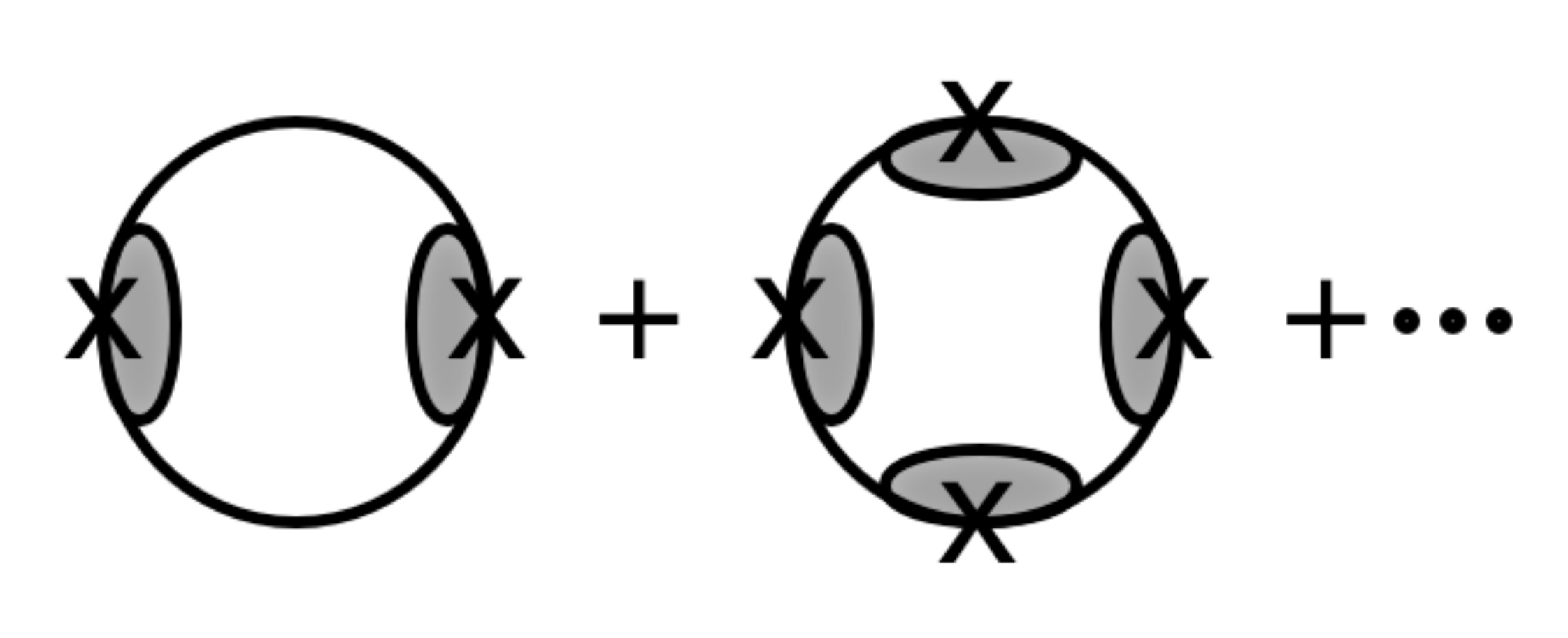,width=3.3in,height=1.3in}
\end{center}
\caption{Vacuum energy density  $\epsilon(m)$ via an infinite summation of massless graphs with zero-momentum dressed $m\tilde{\Gamma}^0_{\rm S}(p,p,0)$ insertions.}
\label{livingwithout2}
\end{figure}

To explicitly look for dynamical symmetry breaking we introduce a trial  mass term that is not present in the action, and rewrite $I^0_{\rm QED}+I_{\rm FF}$ as $I_{\rm MF}+I_{\rm RI}$, where the  mean-field and residual interaction  components are respectively given by
\begin{eqnarray}
I_{\rm MF}&=&I^0_{\rm QED}+\int d^4x\left(-m\bar{\psi}\psi+\frac{m^2}{2g}\right),
\nonumber\\
I_{\rm RI}&=&\int d^4x\left(-\frac{g}{2}\right)\left(\left(\bar{\psi}\psi-\frac{m}{g} \right)^2+\left(\bar{\psi}i\gamma^5\psi\right)^2\right).~~
\label{A20a}
\end{eqnarray}
Neither of the two actions in (\ref{A20a}) is separately chirally symmetric, only their sum is, with $I_{\rm MF}$ acting as the massive fermion $I^m_{\rm QED}$ given in (\ref{A1a}) (the non-chiral-invariant case actually studied in \cite{Johnson1964,Johnson1967,Baker1971a,Adler1971}), to which a cosmological $m^2/2g$ term has been added. As described in \cite{Mannheim1975,Mannheim2016a}, in a volume $V$ the energy-density difference $\epsilon(m)=(\langle\Omega_m|H_{\rm MF}|\Omega_m\rangle-\langle\Omega_0|H_{\rm MF}|\Omega_0\rangle)/V$ between candidate massive and massless fermion vacua $|\Omega_m\rangle$ and $|\Omega_0\rangle$ for the mean-field $I_{\rm MF}$ is given by the infinite sum $\sum(1/n!)G^{(n)}_0(q_{\mu}=0,m=0)m^n$ of dressed massless  fermion graphs with $m\bar{\psi}\psi$ insertions as exhibited in Fig.  (\ref{livingwithout2}). The terms in the sum can be determined analytically since massless graphs are scale invariant at all momenta. On defining
\begin{eqnarray}
\tilde{S}^{-1}_{\mu}(p,m)= \slashed{p}-m\tilde{\Gamma}^0_{\rm S}(p,p,0)=
\slashed{p}-m\left(\frac{-p^2}{\mu^2}\right)^{\frac{\gamma_{\theta}(\alpha)}{2}},~~~~
\label{A21a}
\end{eqnarray}
we find \cite{Mannheim1975} that at $\gamma_{\theta}(\alpha)=-1$ $\epsilon(m)$ is given by
\begin{eqnarray}
\epsilon(m)&=&i\int \frac{d^4p}{(2\pi)^4}\left[{\rm Tr}~{\rm ln}(\tilde{S}^{-1}_{\mu}(p,m))-{\rm Tr}~{\rm ln}(\slashed{p})\right]
\nonumber\\
&=&-\frac{m^2\mu^2}{8\pi^2}\left[{\rm ln}\left(\frac{\Lambda^2}{m\mu}\right)+\frac{1}{2}\right].
\label{A22a}
\end{eqnarray}

\begin{figure}[htpb]
\begin{center}
\includegraphics[width=3.4in,height=1.8in]{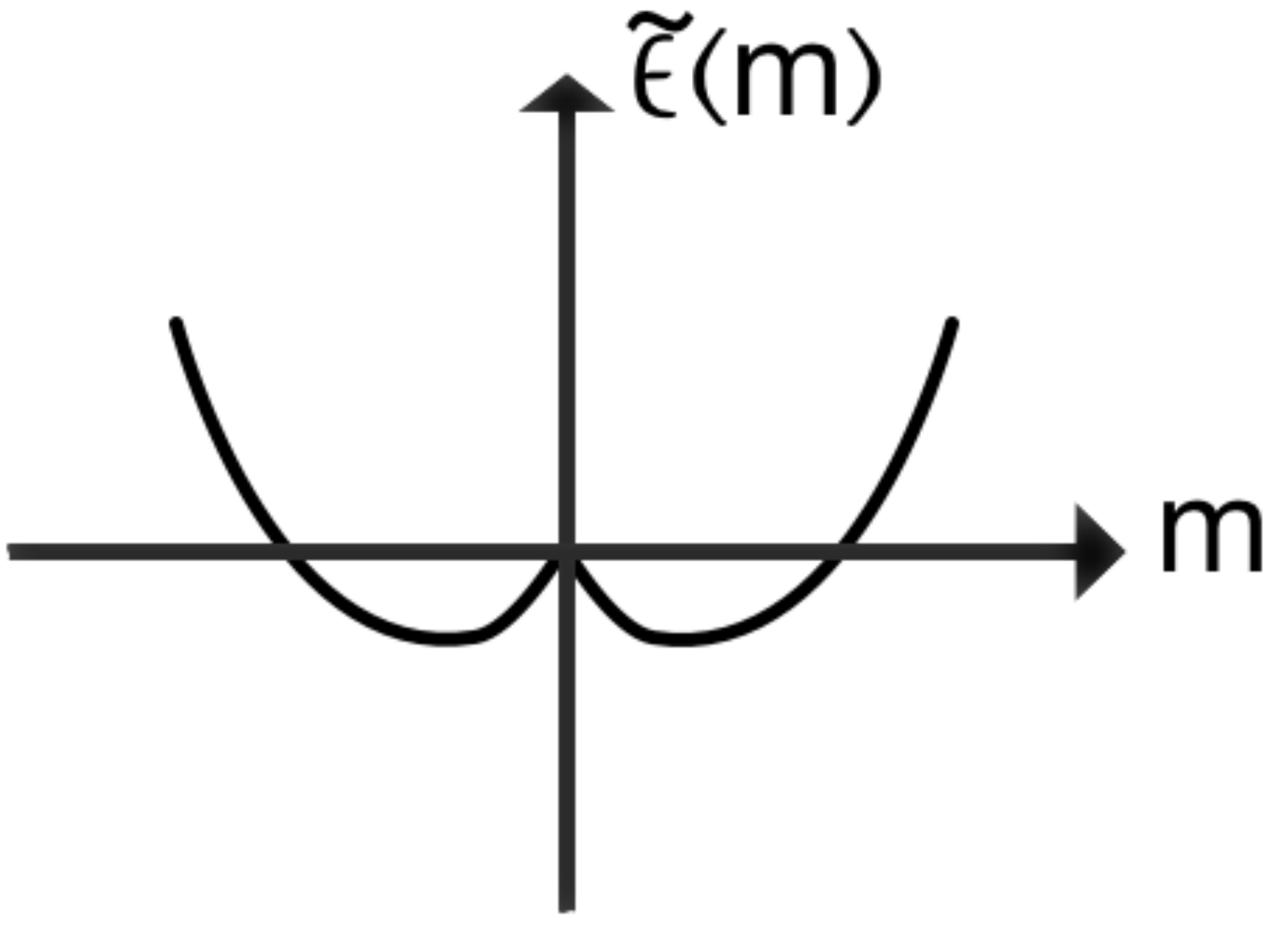}
\end{center}
\caption{$\tilde{\epsilon}(m)$ plotted as a function of  $m$ at $\gamma_{\theta}(\alpha)=-1$.}
\label{livingwithout4}
\end{figure}

By setting 
\begin{eqnarray}
\langle \Omega_M|\left(\bar{\psi}\psi-\frac{M}{g}\right)^2|\Omega_M\rangle&=&
\nonumber\\
\langle \Omega_M|\left(\bar{\psi}\psi-\frac{M}{g}\right)|\Omega_M\rangle^2&=&0
\label{A23a}
\end{eqnarray}
in the mean-field, Hartree-Fock approximation, we define the physical vacuum to be that vacuum in which $\langle\Omega_M|H_{\rm RI}|\Omega_M\rangle=0$ and $\langle\Omega_M|\bar{\psi}\psi|\Omega_M\rangle=M/g$. And with  $\langle \Omega_m|\bar{\psi}\psi|\Omega_m\rangle=\epsilon^{\prime}(m)$ for any $m$, we obtain
\begin{eqnarray}
&&\langle \Omega_M|\bar{\psi}\psi|\Omega_M\rangle=-i\int \frac{d^4p}{(2\pi)^4}{\rm Tr}[\tilde{\Gamma}^0_{\rm S}(p,p,0)\tilde{S}_{\mu}(p,M)]
\nonumber\\
&&=i\int \frac{d^4p}{4\pi^4}\frac{M\mu^2}{(p^2)^2+M^2\mu^2}
=-\frac{M\mu^2}{4\pi^2}{\rm ln}\left(\frac{\Lambda^2}{M\mu}
\right)=\frac{M}{g}.~~~~~
\label{A24a}
\end{eqnarray}
Equation (\ref{A24a}) has a non-trivial solution no matter how small $g$ might be, as long as it is  negative, viz. attractive as per its definition in $I_{\rm FF}$, with symmetry breaking being obtained for weak coupling, and with there being no need for the strong coupling that is thought to be required for dynamical symmetry breaking. Equation (\ref{A24a}) yields the essential singularity behavior familiar from the Bardeen-Cooper-Schrieffer theory of superconductivity, viz. $M\mu=\Lambda^2\exp(4\pi^2/\mu^2g)$. And with (\ref{A24a}) we obtain
\begin{eqnarray}
\tilde{\epsilon}(m)=\epsilon(m)-\frac{m^2}{2g}=\frac{m^2\mu^2}{16\pi^2}\left[{\rm ln}\left(\frac{m^2}{M^2}\right)-1\right],
\label{A25a}
\end{eqnarray}
with the mean-field induced cosmological $-m^2/2g$ term automatically making $\tilde{\epsilon}(m)$ be completely finite without the need for any fine tuning. As seen in Fig. (\ref{livingwithout4}), we recognize $\tilde{\epsilon}(m)$ as having the shape of a double-well potential with a minimum at $m=M$. 

For our purposes here, we note that  $g^{-1}$ of (\ref{A24a}) and  $\Pi^0_{\rm S}(q^2)$ of (\ref{A11a}) both have precisely the same dependence on $\Lambda$. Consequently, $T^0_{\rm S}(q^2)$ and $T^0_{\rm P}(q^2)$ are both automatically finite. Symmetry breaking thus precisely provides the subtraction need to make both $T^0_{\rm S}(q^2)$ and $T^0_{\rm P}(q^2)$ be finite, with the theory doing it automatically all on its own.

\begin{figure}[htpb]
\begin{center}
\includegraphics[width=3.4in,height=0.8in]{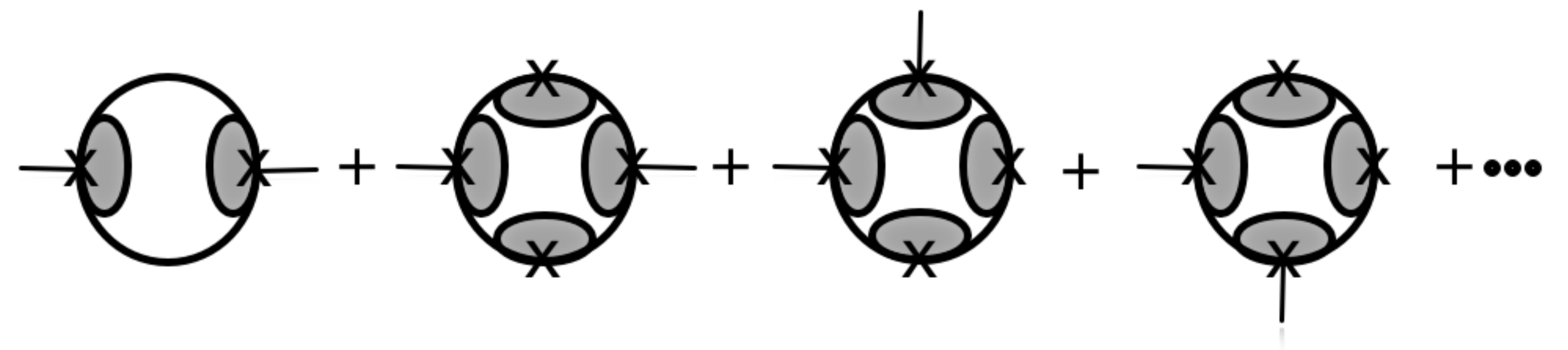}
\end{center}
\caption{$\Pi^m_{\rm S}(q^2)$ developed as an infinite summation of massless graphs, each with two dressed $m\tilde{\Gamma}^0_{\rm S}(p,p+q,q)$ insertions carrying momentum $q_{\mu}$ (shown as external lines), with all other dressed $m\bar{\psi}\psi$  insertions carrying zero momentum.}
\label{livingwithout6}
\end{figure}

The reason why we get an automatic cancellation of ultraviolet divergences is that in the expansion given in Figs.  (\ref{livingwithout2}) and (\ref{livingwithout6}) the only ultraviolet divergent graphs are those with two $\bar{\psi}\psi$ insertions. Thus $\epsilon(m)$, $\Pi_{\rm S}^0(q^2)$ and $\Pi_{\rm P}^0(q^2)$ all have the identical ultraviolet divergence structure, and not just in lowest order but even after being dressed to all orders in $g$ as per Fig. (\ref{higherorder}). The divergent part of $\epsilon(m)$ is given by $(1/2)G^{(2)}_0(q_{\mu}=0,m=0)m^2$, with the divergent part of $\epsilon^{\prime}(m)$ thus being given by $G^{(2)}_0(q_{\mu}=0,m=0)m$. And with $G^{(2)}_0(q_{\mu}=0,m=0)$, $\Pi_{\rm S}^0(q_{\mu})$, and $\Pi_{\rm P}^0(q_{\mu})$ all being identically equal in the massless theory,\footnote{The massless theory $\Pi^0_{\rm S}(q^2)$ can be recognized as the first graph in the summation given in Fig. (\ref{livingwithout6}).} on identifying $\epsilon^{\prime}(m)$ with $m/g$, the cancellation automatically follows.

To reinforce the point,  we look for the Goldstone boson that must be present if there is to be dynamical mass generation since $I^0_{\rm QED}+I_{\rm FF}$ is chirally symmetric. As described in \cite{Mannheim1975,Mannheim1978},  the needed massive theory $\Pi^m_{\rm S}(q^2)$ and  $\Pi^m_{\rm P}(q^2)$ that are to replace the massless $\Pi^0_{\rm S}(q^2)$ and  $\Pi^0_{\rm P}(q^2)$  can be generated via the infinite summation of massless theory graphs given in Fig. (\ref{livingwithout6}), to yield 
\begin{eqnarray}
\Pi_{\rm S}^m(q^2)&=&-i\int \frac{d^4p}{(2\pi)^4}{\rm Tr}\bigg{[}\tilde{S}_{\mu}(p,m)\tilde{\Gamma}^0_{\rm S}(p,p+q,q)
\nonumber\\
&\times&\tilde{S}_{\mu}(p+q,m)\tilde{\Gamma}^0_{\rm S}(p+q,p,-q)\bigg{]},
\nonumber\\
\Pi_{\rm P}^m(q^2)&=&-i\int \frac{d^4p}{(2\pi)^4}{\rm Tr}\bigg{[}\tilde{S}_{\mu}(p,m)\tilde{\Gamma}^0_{\rm S}(p,p+q,q)i\gamma^5
\nonumber\\
&\times&\tilde{S}_{\mu}(p+q,m)\tilde{\Gamma}^0_{\rm S}(p+q,p,-q)i\gamma^5\bigg{]},
\label{A26a}
\end{eqnarray}
where $\tilde{S}_{\mu}(p,m)$ is given in (\ref{A21a}). When $\gamma_{\theta}(\alpha)=-1$, evaluating $\Pi_{\rm P}^m(q^2)$ at $q^2=0$ and $m=M$ gives 
\begin{eqnarray}
\Pi^M_{\rm P}(q^2=0)=i\int \frac{d^4p}{4\pi^4}\frac{\mu^2}{(p^2)^2+M^2\mu^2}.
\label{A27a}
\end{eqnarray}
On comparing with the value of $g^{-1}$ given in (\ref{A24a}), we see that there indeed is a massless pseudoscalar Goldstone boson pole in  $T^M_{\rm P}(q^2)$, just as required. Moreover, since there has to be such a pole if there is to be dynamical mass generation, the log divergences in $1/g$ and in $\Pi^M_{\rm S}(q^2)$ have no choice but to cancel each other identically.  Suppose we now go beyond Hartree-Fock and start dressing Figs. (\ref{livingwithout2}) and (\ref{livingwithout6}) with higher order $I_{\rm FF}$ interactions, just like those in Fig. (\ref{higherorder}). The Goldstone pole must survive, and thus the cancellation of the log divergence must persist, with the all order in $g$ dressings of $\epsilon^{\prime}(M)=\langle \Omega_M|\bar{\psi}\psi|\Omega_M\rangle=M/g$ and of $T^M_{\rm P}(q^2=0)$ both automatically yielding the previously found condition $4\pi^2G^{-1}=-F(G){\rm ln}[\Lambda^2/M\mu]$. Thus a condition that was initially imposed in order to control the ultraviolet behavior of $I^0_{\rm QED}+I_{\rm FF}$  is now found to automatically emerge as a constraint provided by the consistency of the infrared structure of the theory. 

Because of the chiral symmetry, in the same way that the massive $T^M_{\rm P}(q^2)$ contains a bound state Goldstone pole, the massive $T^M_{\rm S}(q^2)$ contains a dynamical scalar Higgs boson. Interestingly, explicit calculation \cite{Mannheim2015,Mannheim2016,Mannheim2016a} shows that it actually lies above the threshold in the scalar channel fermion-antifermion scattering amplitude, to thus be a resonance with a width, rather than a bound state. In \cite{Mannheim2015,Mannheim2016,Mannheim2016a} it was suggested that this width could serve as a diagnostic to distinguish between a dynamical Higgs boson and the elementary one that would appear in a  fundamental Lagrangian.

With $I_{\rm FF}$ having become renormalizable, our work can be viewed as a renormalizable version of the Nambu-Jona-Lasinio model \cite{Nambu1961}, with the point vertices of that model having been softened by $\gamma_{\theta}(\alpha)=-1$ just enough to make them renormalizable. Moreover, not only do we obtain renormalizability, with $g$ being fixed via the Hartree-Fock (\ref{A24a}) we actually obtain finiteness, with $T^M_{\rm S}(q^2)$ and $T^M_{\rm P}(q^2)$ both automatically being finite.\footnote{Our finite, cutoff-independent, result should be contrasted with prior studies of QED coupled to a four-fermion interaction since those studies all involved a cutoff for the four-fermion interaction (see e.g. \cite{Miransky1993,Mannheim2015,Mannheim2016a}). The reason why our result differs from prior results is that while all studies use the same quenched ladder approximation type fermion propagator $S_m^{-1}(p)$ as given in (\ref{A3a}), unlike these other studies we use the dressed  $\tilde{\Gamma}^0_{\rm S}(p,p+q,q)=[(-p^2)/\mu^2)(-(p+q)^2)/\mu^2)]^{\gamma_{\theta}(\alpha)/4}$ vertex that is given in (\ref{A8a}), while other studies (see e.g. \cite{Leung1986}) took this vertex to be the pointlike $\tilde{\Gamma}^0_{\rm S}(p,p+q,q)=1$, and thus needed a cutoff since the asymptotic behavior was not convergent enough.}

While we have only studied four-fermion interactions in flat spacetime in the present paper, having a renormalizable $I_{\rm FF}$ is also advantageous for the treatment of the vacuum energy density in the presence of gravity. Specifically, since gravity couples to energy density itself and not to energy density difference, in the presence of gravity one cannot normal order away vacuum energy density infinities. Rather, one must cancel them dynamically, and this can be  achieved by $I_{\rm FF}$ if $\gamma_{\theta}(\alpha)=-1$ \cite{Mannheim2015,Mannheim2016,Mannheim2016a}. In \cite{Mannheim2015,Mannheim2016,Mannheim2016a} it is also shown how a renormalizable $I_{\rm FF}$ and a gravity theory that is conformal can control the cosmological constant term that is induced when mass is dynamically generated. A renormalizable $I_{\rm FF}$ thus plays a dual role, participating in dynamical symmetry breaking and in canceling infinities in the vacuum energy density.

\end{document}